\documentclass[10pt,a4paper]{article}

\usepackage[T1]{fontenc}
\usepackage[utf8]{inputenc}

\usepackage{libertinus}
\usepackage{microtype}

\usepackage{graphicx}
\usepackage{amsmath}
\usepackage{natbib}

\usepackage{microtype}
\usepackage[a4paper,top=15mm,bottom=16mm,left=17mm,right=17mm]{geometry}

\usepackage{booktabs}
\usepackage{array}
\usepackage{tabularx}
\usepackage{multirow}
\usepackage{amssymb}

\usepackage[table,dvipsnames]{xcolor}

\usepackage{enumitem}
\usepackage{titlesec}
\usepackage{fancyhdr}
\usepackage{lastpage}
\usepackage{hyperref}
\usepackage{url}

\usepackage{subcaption}

\definecolor{AtlanticBlue}{HTML}{164A63}
\definecolor{OceanBlue}{HTML}{28799E}
\definecolor{MistBlue}{HTML}{EAF3F7}
\definecolor{SlateGrey}{HTML}{4B5963}
\definecolor{WarmGrey}{HTML}{F4F4F2}
\definecolor{RuleGrey}{HTML}{CBD3D7}

\hypersetup{
  colorlinks=true,
  linkcolor=AtlanticBlue,
  citecolor=AtlanticBlue,
  urlcolor=OceanBlue,
  pdftitle={Assessing Flood-Related Wellbeing from Public Discourse},
  pdfauthor={Roisin Luo and Karyn Morrissey}
}

\renewcommand{\arraystretch}{1.14}
\setlist[itemize]{leftmargin=1.4em,itemsep=1pt,topsep=2pt}

\titleformat{\section}
  {\large\bfseries\color{AtlanticBlue}}
  {\thesection.}{0.45em}{}
  [\vspace{-2pt}\color{RuleGrey}\titlerule]

\titlespacing*{\section}{0pt}{8pt}{4pt}

\titleformat{\subsection}
  {\normalsize\bfseries\color{SlateGrey}}
  {\thesubsection.}{0.4em}{}

\titlespacing*{\subsection}{0pt}{6pt}{2pt}

\newcolumntype{L}[1]{>{\raggedright\arraybackslash}p{#1}}
\newcolumntype{C}[1]{>{\centering\arraybackslash}p{#1}}
\newcolumntype{Y}{>{\raggedright\arraybackslash}X}

\begin{document}

\begin{center}
  {\color{AtlanticBlue}\rule{\textwidth}{1.2pt}}\\[6pt]


  {\LARGE\bfseries\color{AtlanticBlue}
  A population-level assessment framework for flood-related wellbeing from public discourse in Ireland}

  \vspace{7pt}

  {\large R\'ois\'in Luo\footnotemark[1] \quad and \quad Karyn Morrissey}

  \vspace{3pt}

  {\normalsize
  J.E. Cairnes School of Business \& Economics,
  University of Galway}

  \vspace{2pt}


  {\color{AtlanticBlue}\rule{\textwidth}{0.5pt}}
\end{center}

\footnotetext[1]{Corresponding author: \href{mailto:jiaolin.luo@universityofgalway.ie}{\nolinkurl{jiaolin.luo@universityofgalway.ie}}}

\noindent
\fcolorbox{RuleGrey}{MistBlue}{
\begin{minipage}{0.965\textwidth}
\small
\textbf{\color{AtlanticBlue}Abstract.}

\paragraph{Background.}
Ireland's position on the eastern North Atlantic exposes it to moisture-laden westerlies and frequent low-pressure systems, generating abundant precipitation and substantial pluvial and fluvial flood risk. Climate change is intensifying heavy rainfall and compound flood risk, with consequences extending beyond physical damage and conventional economic indicators. 

\paragraph{Method.}
We present a framework for assessing flood-related wellbeing from unobtrusive public discourse over time. The framework develops an assessment instrument comprising three complementary constructs: \emph{distress}, \emph{functional disruption}, and \emph{institutional alienation}, capturing flood-related impacts on affective appraisal, daily functioning, and social and institutional connectedness. To enable population-scale analysis of complex cognitive and psychological responses, we develop a dedicated flood-related wellbeing reasoning model (\textsc{Wellbeing-Former}) that reads evidence of flood-related wellbeing, assigns evidence-grounded scores, and produces transparent rationales.

\paragraph{Data, Results \& Implications.}
Using the research platform \textsc{MCL} (Meta Content Library), we construct a flood-related discourse dataset by querying posts from Ireland containing the terms \emph{flood}, \emph{rain}, and \emph{storm}. The resulting dataset comprises approximately \textbf{$224{,}000$} posts produced between 2012 and 2016. We apply the resulting \textsc{Wellbeing-Former} to this dataset to conduct a population-level analysis. This approach provides temporally sensitive evidence of experienced and anticipated flood impacts, complements survey-based assessment, and supports a multidimensional understanding of wellbeing under climate-related hazards for policy development, adaptation planning, public communication, and decision-making.

\end{minipage}
}

\vspace{3pt}


\section{Introduction}

Ireland's position on the eastern margin of the North Atlantic exposes the country to prevailing westerly airflow and frequent Atlantic low-pressure systems, which transport moisture-laden maritime air and generate abundant precipitation, thereby contributing to substantial pluvial and fluvial flood risk. Climate change is altering Ireland's hydrological cycle and precipitation regime, with projected increases in heavy rainfall and compound events involving rainfall, river flow, storm surge, and high tides \cite{epa2023,met2024}. The assessment of flood-related wellbeing is therefore essential for characterising impacts that are not adequately captured by measures of physical damage and aggregate economic loss. Floods and severe weather can give rise to affective distress and uncertainty, disrupt mobility and access to essential services, damage homes and livelihoods, and undermine confidence in institutions responsible for prevention, response, and recovery.

Wellbeing impacts are conventionally assessed using self-reported surveys, questionnaires, and interviews. Although these approaches provide valuable evidence, they can be expensive, episodic, geographically limited, and susceptible to recall and response biases. Social media platforms offer an alternative source of unobtrusive public discourse at scale; however, these data are unstructured, fragmented, and context-dependent, making the reliable identification and interpretation of flood-related wellbeing evidence challenging.

Extracting flood-related wellbeing evidence from public discourse can provide a large and continuously generated source of data. However, two challenges impede its effective use. First, dedicated and reliable instruments for unobtrusive wellbeing assessment remain underdeveloped. Second, a fundamental trade-off exists between interpretability and computational efficiency. Lexicon-based methods are transparent and efficient but often fail to capture context-dependent cognitive appraisals and complex affective responses. In contrast, large language models (LLMs) can approach human-level performance on some psychological reasoning tasks and produce interpretable, rationale-grounded assessments; however, reasoning with state-of-the-art LLMs at population scale remains computationally expensive.

These gaps motivate this research, which makes the following contributions:
\begin{itemize}
    \item \textbf{Wellbeing Instrument for Flood-Related Unobtrusive Discourse.} We develop a dedicated assessment instrument for measuring flood-related wellbeing from unobtrusive public discourse. The instrument comprises three components: (i) three constructs measuring \emph{distress}, \emph{functional disruption}, and \emph{institutional alienation}; (ii) scales and scoring rubrics specifying four evidence levels: 0 (\emph{zero evidence}), 1 (\emph{low evidence}), 2 (\emph{moderate evidence}), and 3 (\emph{high evidence}); and (iii) an evidence--construct assignment matrix.

    \item \textbf{Interpretable and Efficient Wellbeing Reasoning Model.} We design a dedicated rationale-based Bayesian reasoning model, \textsc{Wellbeing-Former}, designed to trade-off among transparency, computational efficiency, and predictive performance. This model establishes state-of-the-art performance for flood-related wellbeing assessment against strong baselines, including lexicon-based methods, neural language modeling, selective rationales, and a frontier LLM.

    \item \textbf{Population-Level Analysis of Flood-Related Wellbeing in Ireland.} We apply the assessment instrument and \textsc{Wellbeing-Former} to approximately $120{,}000$ public posts concerning Ireland produced between 2012 and 2016, providing a population-level temporal analysis of experienced and anticipated flood-related wellbeing. The resulting analysis can inform stakeholders involved in policy development, adaptation planning, public communication, and decision-making.

\end{itemize}

\section{Flood-Related Wellbeing Instrument}
\label{sec:instrument}

Rater-based assessment of wellbeing with unobtrusive public discourse is constrained by the absence of content producers' daily wellbeing baselines and the unreliability of inferring their mental states through subjective empathy. We therefore introduce an objective, evidence-based scheme for assessing wellbeing as articulated in the discourse, comprising three components: construct definitions, ordinal scales and scoring rubrics, and an evidence--construct assignment matrix.

\paragraph{Wellbeing Instrument.}
The instrument comprises three complementary negative-pole constructs. \emph{Distress} represents flood-related affective and cognitive suffering associated with actual or anticipated flooding and its consequences. \emph{Functional disruption} represents flood-related impairment of everyday functioning, material stability, and the performance of essential domestic, occupational, educational, caregiving, and community roles. \emph{Alienation} represents flood-related social or institutional disconnection, including perceived exclusion, abandonment, neglect, or failure by interpersonal networks, authorities, insurers, or public institutions. The three constructs respectively operationalise the psychological, functional, and relational dimensions of flood-related wellbeing.

\paragraph{Scales.}
Each construct is assessed independently using a four-level ordinal scale: 0 indicates \emph{no evidence}, 1 indicates \emph{low evidence}, 2 indicates \emph{moderate evidence}, and 3 indicates \emph{high evidence}. These levels represent the strength, clarity, intensity, extent, and centrality of the evidence contained in the discourse; they do not represent the magnitude of change from the content producer's unobserved personal baseline. The assigned score is the highest evidence level supported by explicit textual cues or by consequences directly and reasonably implied by the discourse. In the absence of admissible evidence, the construct is assigned a score of 0.

\paragraph{Evidence--Construct Assignment Matrix.}
Evidence is organised into two categories: \emph{stressor evidence} and \emph{appraisal \& response evidence}. Stressor evidence identifies the flood-related hazard or harm that may precipitate a wellbeing impact, whereas appraisal \& response evidence identifies the psychological, functional, or relational manifestation of that impact. A single item of evidence may be relevant to more than one construct, while other evidence is construct-specific. The evidence--construct assignment matrix formally specifies these admissible correspondences. Its purpose is to permit theoretically justified overlap while preventing evidence associated with one construct from being used to support another without an explicit conceptual basis. The matrix thereby constrains cue retrieval, reduces construct leakage, and ensures that each ordinal score is traceable to observable and construct-relevant evidence.

\begin{figure}
    \centering
    \includegraphics[width=0.9\linewidth]{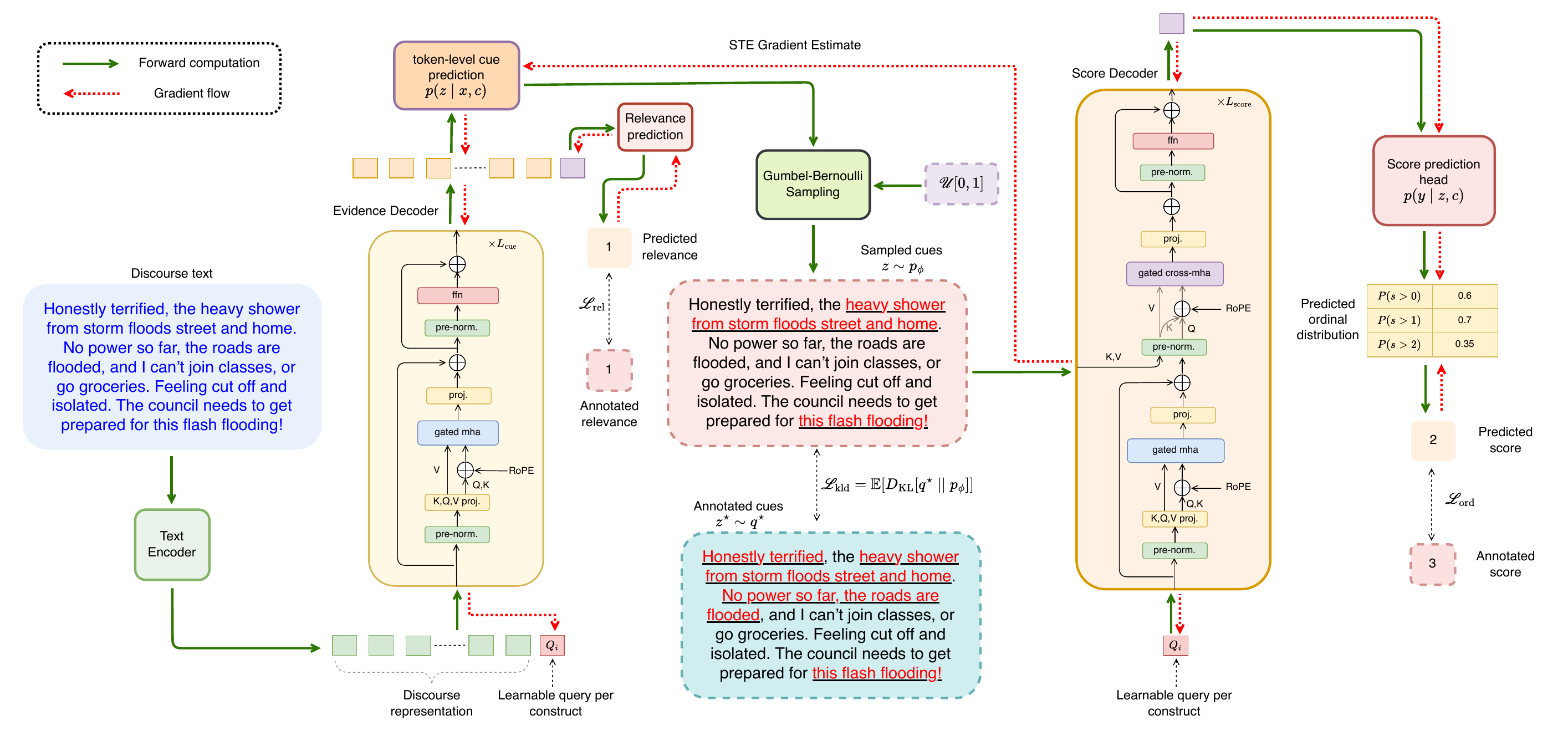}
    
    \caption{\textbf{Wellbeing-Former Architecture.} \textsc{Wellbeing-Former} comprises three components: (i) a shared lightweight pre-trained language encoder; (ii) a non-causal evidence decoder that predicts a factorized Bernoulli distribution over token-level cue selections; and (iii) a non-causal score decoder that uses representations selected by the sampled cue masks as cross-attention \emph{keys} and \emph{values} for rationale-based ordinal scoring.}
    
    \label{fig:architecture_wellbeingformer}
    
\end{figure}

\section{Data Collection}

Using the research platform \textsc{MCL}\footnote{In accordance with the GDPR and Meta's policy, access to MCL is restricted to screened researchers.} (Meta Content Library)~\citep{Meta_MCL_v6}, we create a flood-related discourse dataset by querying for the terms \emph{flood}, \emph{rain}, and \emph{storm} in Ireland. The resulting dataset, termed \emph{Flood-Related Irish Discourse}, comprises $224{,}000$ posts. We randomly sample $10{,}000$ posts as train set and employ state-of-the-art Anthropic's Claude Opus 4.8 as the LLM rater for corpus annotation.

\section{Method}

\paragraph{Reasoning Model.}
Figure~\ref{fig:architecture_wellbeingformer} presents the architecture of \textsc{Wellbeing-Former}, which comprises three components: (i) a shared lightweight pre-trained language encoder; (ii) a non-causal evidence decoder that predicts a factorised Bernoulli distribution over token-level cue selections; and (iii) a non-causal score decoder that uses the representations selected by the sampled cue masks as cross-attention \emph{keys} and \emph{values} for rationale-based ordinal scoring. We train \textsc{Wellbeing-Former} on the annotated \emph{Flood-Related Irish Discourse} dataset and apply the resulting model to the full corpus of $224{,}000$ posts to extract flood-related wellbeing scores.

\paragraph{Score Standardisation \& Aggregation.}
Let $c_i=(c_i^{(j)})_{j=1}^{n}$ denote the raw series for the $i$-th construct, with mean $\mu(c_i)$ and standard deviation $\sigma(c_i)$. We standardise the scores as follows:
\begin{align}
    \widetilde{c}_i = \frac{c_i - \mu(c_i)}{\sigma(c_i)}.
\end{align}
To aggregate the measured scores into a scalar wellbeing score, we apply an ordinality-preserving, strictly monotonic functional:
\begin{align}
    y = \sigma(-\beta \sum_i \widetilde{c}_i), \qquad \sigma(x) = \frac{1}{1 + \mathrm{e}^{-x}},
\end{align}
where $\beta$ is referred to as the \emph{aggregation sensitivity} coefficient.



\begin{table*}[t]
\centering

\caption{\textbf{Strong-Baseline Comparison on Irish Flood-related Discourse.} Given the inherent uncertainty of wellbeing assessment across raters, we adopt a within-one-level agreement protocol, under which predicted and annotated scores differing by at most one ordinal level are considered to agree. We report accuracy ($\operatorname{Acc}\text{@}1$), $F_1\text{@}1$, Cohen's kappa ($\kappa\text{@}1$), and Kendall's rank correlation ($\tau\text{@}1$) under this protocol. Rationale retrieval is evaluated using token-level accuracy, $F_1$, and recall.}


\label{tab:benchmark}

\begingroup

\definecolor{sentimentrow}{RGB}{255,251,235}
\definecolor{emotionrow}{RGB}{240,253,244}
\definecolor{lexiconrow}{RGB}{239,246,255}

\definecolor{neuralrow}{RGB}{245,243,255}
\definecolor{rationalerow}{RGB}{236,254,255}

\definecolor{llmrow}{RGB}{255,241,242}

\definecolor{oursrownorl}{RGB}{255,241,242}
\definecolor{oursrowrl}{RGB}{255,215,219}

\setlength{\tabcolsep}{5pt}
\renewcommand{\arraystretch}{1.10}

\resizebox{0.9\linewidth}{!}{
\begin{tabular}{@{}lcccccccc@{}}
\toprule
& \multicolumn{4}{c}{\textbf{Score Assignment}}
& \multicolumn{3}{c}{\textbf{Rationale Retrieval}}
& \\
\cmidrule(lr){2-5}
\cmidrule(lr){6-8}

\textbf{Method}
& $\operatorname{Acc}\text{@}1$
& $F_1\text{@}1$
& $\kappa\text{@}1$
& $\tau\text{@}1$
& \textbf{Token Acc.}
& \textbf{$F_1$}
& \textbf{Recall}
& \textbf{Interpretable} \\
\midrule

\multicolumn{9}{@{}l}{\textit{Lexicon-Based Methods}} \\

\rowcolor{lexiconrow}
VADER~\citep{Hutto_Gilbert_2014}
& 0.811
& 0.498
& 0.493
& 0.285
& --
& --
& --
& $\checkmark$ \\

\rowcolor{lexiconrow}
AFINN~\citep{nielsen2011new}
& 0.837
& 0.581
& 0.565
& 0.393
& --
& --
& --
& $\checkmark$ \\

\rowcolor{lexiconrow}
Opinion Lexicon~\citep{hu2004mining}
& 0.814
& 0.508
& 0.501
& 0.297
& --
& --
& --
& $\checkmark$ \\

\rowcolor{lexiconrow}
SentiWordNet~\citep{baccianella2010sentiwordnet}
& 0.811
& 0.494
& 0.491
& 0.282
& --
& --
& --
& $\checkmark$ \\

\rowcolor{lexiconrow}
ANEW / Warriner VAD~\citep{bradley1999anew,warriner2013norms}
& 0.826
& 0.552
& 0.536
& 0.359
& --
& --
& --
& $\checkmark$ \\

\rowcolor{lexiconrow}
labMT / Hedonometer~\citep{dodds2011temporal}
& 0.814
& 0.506
& 0.500
& 0.296
& --
& --
& --
& $\checkmark$ \\


\rowcolor{lexiconrow}
NRC Emotion (EmoLex)~\citep{mohammad2013crowdsourcing}
& 0.858
& 0.602
& 0.616
& 0.502
& --
& --
& --
& $\checkmark$ \\

\rowcolor{lexiconrow}
LIWC~\citep{pennebaker2015liwc}
& 0.892
& 0.691
& 0.711
& 0.624
& --
& --
& --
& $\checkmark$ \\

\rowcolor{lexiconrow}
Empath~\citep{fast2016empath}
& 0.833
& 0.553
& 0.546
& 0.392
& --
& --
& --
& $\checkmark$ \\



\addlinespace
\multicolumn{9}{@{}l}{\textit{Neural Language Modeling}} \\

\rowcolor{neuralrow}
BiLSTM + Linear
& $0.961 \pm 0.011$
& $0.821 \pm 0.023$
& $0.899 \pm 0.030$
& $0.876 \pm 0.036$
& --
& --
& --
& -- \\

\rowcolor{neuralrow}
BiGRU + Linear~\citep{cho2014learning}
& $0.957 \pm 0.022$
& $0.806 \pm 0.052$
& $0.886 \pm 0.058$
& $0.864 \pm 0.065$
& --
& --
& --
& -- \\

\rowcolor{neuralrow}
TextCNN + Linear~\citep{kim2014convolutional}
& $0.918 \pm 0.027$
& $0.714 \pm 0.069$
& $0.778 \pm 0.075$
& $0.731 \pm 0.087$
& --
& --
& --
& -- \\

\rowcolor{neuralrow}
BERT + Linear
& $0.931 \pm 0.011$
& $0.766 \pm 0.019$
& $0.816 \pm 0.029$
& $0.767 \pm 0.037$
& --
& --
& --
& -- \\

\addlinespace
\multicolumn{9}{@{}l}{\textit{Selective Rationalization}} \\

\rowcolor{rationalerow}
REFER~\citep{ghasemi2023refer}
& $0.970 \pm 0.002$
& $0.845 \pm 0.003$
& $0.922 \pm 0.004$
& $0.896 \pm 0.005$
& $0.693 \pm 0.001$
& $0.375 \pm 0.002$
& $0.537 \pm 0.004$
& $\checkmark$ \\

\rowcolor{rationalerow}
RTP~\citep{brinner2024rationalizing}
& $0.966 \pm 0.022$
& $0.852 \pm 0.066$
& $0.910 \pm 0.065$
& $0.884 \pm 0.081$
& $0.693 \pm 0.004$
& $0.378 \pm 0.004$
& $0.536 \pm 0.009$
& $\checkmark$ \\

\rowcolor{rationalerow}
N2R~\citep{liu2025breaking}
& $0.934 \pm 0.046$
& $0.775 \pm 0.085$
& $0.825 \pm 0.118$
& $0.770 \pm 0.156$
& $0.666 \pm 0.001$
& $0.308 \pm 0.008$
& $0.411 \pm 0.005$
& $\checkmark$ \\

\addlinespace
\multicolumn{9}{@{}l}{\textit{LLM Rater Baseline}} \\

\rowcolor{llmrow}
ChatGPT-5.6 Sol High
& 0.868
& 0.713
& 0.649
& 0.522
& 0.675
& 0.135
& 0.206
& $\checkmark$ \\

\midrule
\multicolumn{9}{@{}l}{\textit{Ours}} \\

\rowcolor{oursrowrl}
\textsc{Wellbeing-Former}
& \textcolor{blue}{$\mathbf{0.982 \pm 0.004}$}
& \textcolor{blue}{$\mathbf{0.918 \pm 0.016}$}
& \textcolor{blue}{$\mathbf{0.953 \pm 0.013}$}
& \textcolor{blue}{$\mathbf{0.946 \pm 0.013}$}
& \textcolor{blue}{$\mathbf{0.814 \pm 0.021}$}
& $\mathbf{0.378 \pm 0.006}$
& $0.494 \pm 0.046$
& $\checkmark$ \\

\bottomrule
\end{tabular}
}

\endgroup

\vspace{-1em}

\end{table*}

\begin{figure}[t]
    \centering

    \begin{subfigure}[t]{0.85\linewidth}
        \centering
        \includegraphics[width=\linewidth]{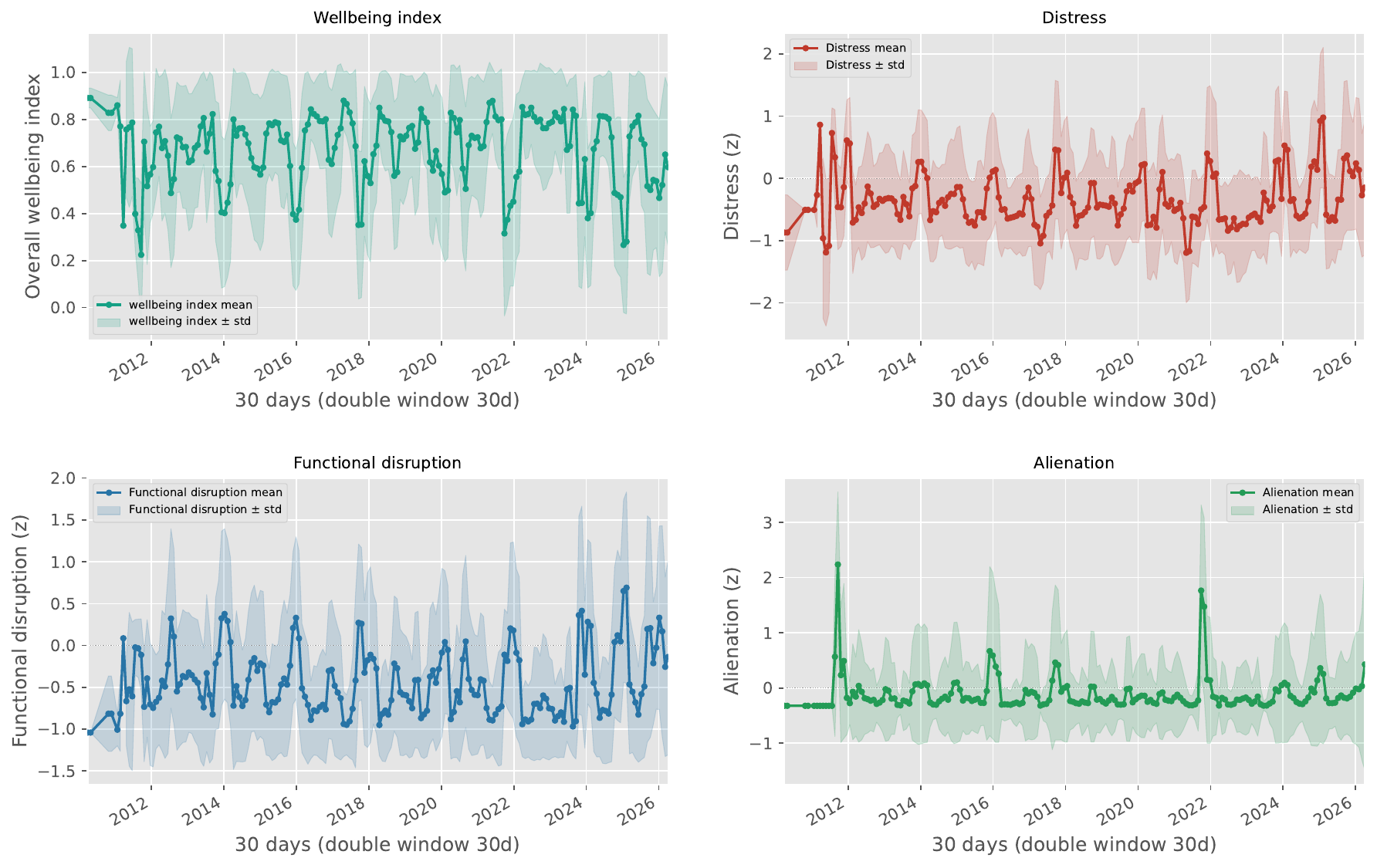}
        
        \caption{Temporal analysis of standarised scores for wellbeing constructs and the aggregated score.}
        
        \label{fig:wellbeing-temporal}
    \end{subfigure}

    \vspace{4pt}

    \begin{subfigure}[t]{0.85\linewidth}
        \centering
        \includegraphics[width=\linewidth]{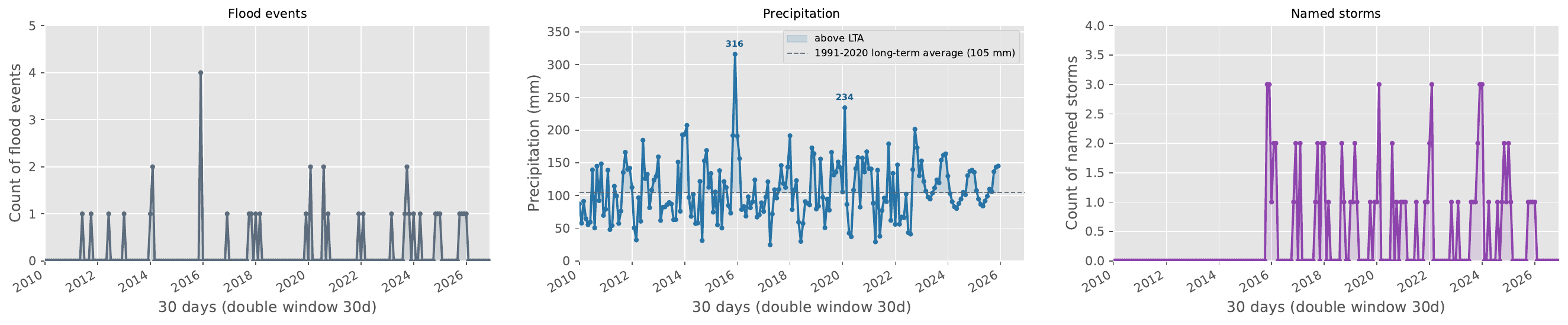}
        
        \caption{Climate reference.}
        
        \label{fig:climate-reference}
    \end{subfigure}

    \caption{\textbf{Temporal relationship between flood-related wellbeing and climate indicators.} Panel (a) presents the standardised flood-related wellbeing series, while Panel (b) presents the corresponding climate-reference series. The positive wellbeing index is standarized by aggregating the three construct scores through the monotonic, ordinality-preserving functional $y=\sigma[-\beta(c_1+c_2+c_3)]$, scaling wellbeing index to $(0, 1)$, where $\sigma$ denotes the sigmoid function, $\beta$ is a sensitivity coefficient set to $1$ by default, and $c_i$ denotes the $i$-th $z$-normalized construct score. Consequently, higher flood-related impact scores produce lower wellbeing-index values.}
    
    \label{fig:temporal-wellbeing-climate}
\end{figure}

    
    

\section{Results \& Findings}

\paragraph{Strong-Baseline Benchmarking.} We organize the baselines into four groups: \emph{lexicon-based methods}, \emph{neural language models}, \emph{selective rationalization}, and an \emph{LLM rater baseline} as reference. As reported in Table~\ref{tab:benchmark}, averaged over four runs, \textsc{Wellbeing-Former} establishes state-of-the-art performance across all five groups, attaining an $\operatorname{Acc}\text{@}1$ of $0.982$, an $F_1\text{@}1$ of $0.918$, a Cohen's kappa ($\kappa\text{@}1$) of $0.953$, and a Kendall's rank correlation ($\tau\text{@}1$) of $0.946$. It also achieves a token-level rationale-retrieval accuracy of $0.814$, and a $F_1\text{@}1$ of $0.378$, demonstrating reliable construct-relevant evidence retrieval and ordinal score assignment.

\paragraph{Population-Level Analysis in Ireland.} Figure~\ref{fig:temporal-wellbeing-climate} shows the standarized construct scores and the aggregated wellbeing index with sensitivity coefficient $\beta=1$ spanning 2012--2026, and the climate reference including flood frequency, precipitation and named storms among the same period. We then use Bayesian state-space model for studying the correlation between flood and wellbeing. The modeling show that a one standard-deviation increase in flood occurrence was associated with a cumulative increase of $0.970$ standard deviations in the wellbeing index over the contemporaneous period and three subsequent periods. The model estimates with $95\%$ posterior probability that this cumulative increase lies between $0.217$ and $1.984$ standard deviations. The estimated increase in the contemporaneous period was $0.157$ standard deviations, with a $95\%$ posterior range from $-0.087$ to $0.479$, while the largest increase occurred one period after flood occurrence.

\section{Conclusion}
This study presents a population-level assessment framework for flood-related wellbeing from unobtrusive public discourse. We first develop an evidence-based flood-wellbeing instrument for analyzing flood-related discourse, consisting of three components: wellbeing constructs, ordinal scales, and an evidence--construct assignment matrix. To automate the analysis at the population level, we then design a dedicated reasoning model, \textsc{Wellbeing-Former}. Finally, we apply \textsc{Wellbeing-Former} to the large-scale \emph{Flood-Related Irish Discourse} corpus comprising $224{,}000$ posts. Our findings reveal that floods have non-negligible impacts on \emph{distress}, \emph{functional disruption}, and \emph{institutional alleviation}. Overall, our framework provides temporally sensitive evidence of both experienced and anticipated flood impacts, complements survey-based assessments, and supports a multidimensional understanding of wellbeing under climate-related hazards for policy development, adaptation planning, public communication, and decision-making.

\section*{Acknowledgements}

This work was conducted as part of Project Digi\'Eire at the J.E. Cairnes School of Business \& Economics, University of Galway.

\bibliographystyle{plain}
\bibliography{references}

\end{document}